\documentstyle[11pt,newpasp,twoside,epsf,graphicx]{article}
\def\edcomment#1{\iffalse\marginpar{\raggedright\sl#1\/}\else\relax\fi}
\reversemarginpar

\begin{document}
\title{The Role of Dust in Producing the Cosmic Infrared Background}
\author{Eli Dwek}
\affil{Laboratory for Astronomy and Solar Physics, NASA Goddard Space Flight
Center, Greenbelt, Maryland 20771, USA}

\begin{abstract}
The extragalactic background light (EBL), exclusive of the cosmic microwave
background, consists of the cumulative radiative output from
all energy sources in the universe since the epoch of
recombination. Most of this energy is released at ultraviolet
and optical wavelengths. However, observations show that a
significant fraction of the EBL falls in the 10 to 1000~$\mu$m
wavelength regime. This provides conclusive evidence that we
live in a dusty universe, since only dust can efficiently
absorbs a significant fraction of the background energy and reemit it at
infrared wavelengths. The general role of dust in forming the
cosmic infrared background (CIB) is therefore obvious.
However, its role in determining the exact spectral shape of
the CIB is quite complex. The CIB spectrum depends on the
microscopic physical properties of the dust, its composition,
abundance, and spatial distribution relative to the emitting
sources, and its response to evolutionary processes that can
modify all the factors listed above. This paper will present a
brief summary of the many ways dust affects the intensity and
spectral shape of the cosmic infrared background.  In an Appendix 
we present new limits on the mid-infrared intensity 
of the CIB using TeV $\gamma$-ray observations of Mrk 501.
\end{abstract}

\section{Introduction}
The extragalactic background light (EBL), exclusive of the
cosmic microwave background, is the repository of all radiant energy
releases in the universe since the
epoch of recombination. Radiative sources contributing to the EBL include
stars, which derive their energy
from the nuclear processing of hydrogen into heavier elements, active
galactic nuclei (AGN), which are
powered by the release of gravitational energy associated with the accretion
of matter onto a central
black hole, and various exotic sources, such as decaying particles,
primordial black holes, exploding
stars, and substellar mass objects. Current limits on the EBL, and the
relative contribution of the
various energy sources to the EBL are presented in these  Symposium
Proceedings by Hauser (2001) and in
the review paper of Hauser \& Dwek (2001).

In a dust-free universe the EBL can, in principle, be simply
derived from knowledge of the spectrum of the emitting sources and the
cosmic history of
their energy release. In a dusty universe the total intensity of the EBL is
unchanged, but its
energy is redistributed over the entire X-ray to far-infrared region of the
spectrum. Predicting the EBL
spectrum in a dusty universe therefore poses a significant challenge, since
the exact frequency
distribution of the reradiated emission depends on a large number of
factors. On a microscopic level, the
emitted spectrum depends on the wavelength dependence of the absorption and
scattering properties of the
dust, which in turn depend on the dust composition and size distribution.
The reradiated spectrum also
depends on the dust abundance and the relative spatial distribution of
energy sources and absorbing dust.
Finally, the cumulative spectrum from all sources depends on various
evolutionary factors, including the
history of dust formation and processes which destroy the dust, modify it,
or re-distribute it relative to
the radiant sources. Intergalactic dust, if present in sufficient
quantities, can cause an overall
dimming of the UV-optical output from distant sources, and produce a truly
diffuse infrared
background. In steady-state models for the universe, dust plays a more
significant role,
producing the cosmic microwave background via the thermalization of
starlight by iron
whiskers (Hoyle, Burbidge,
\& Narlikar 1993). In the following I will examine in more detail the
various factors and processes that
determine the intensity and spectral energy distribution (SED) of the cosmic
infrared background (CIB).

\section{Dust Properties}

The presence of dust in the interstellar medium (ISM) of the Milky Way is
manifested in many
different ways including the extinction, scattering, and polarization of
starlight, the infrared emission,
the interstellar depletion, and the presence of isotopic anomalies in
meteorites. The most accepted interstellar dust model consists of a
population of bare
silicate and graphite grains with a power law distribution in grain radii
extending from a few tens of
angstroms to
 about 0.5~$\mu$m. An additional population of macromolecules, most commonly
identified with
polycyclic aromatic hydrocarbons (PAHs) must be added to this model.
The very small dust particles and macromolecules are stochastically heated
by the ambient radiation field, and give rise to the mid-infrared continuum
spectrum and mid-infrared emission features observed in the Milky Way and in
external galaxies.

Figure~1 (left panel) presents the mass absorption coefficient of graphite
and
silicate dust grains as a function of wavelength. The
Figure illustrates the different absorption efficiencies of carbonaceous and
silicate dust
particles, the latter being significantly more transparent in the UV-visible
regions of the spectrum.
The efficiency of converting starlight to thermal infrared emission clearly
depends on the relative
abundance of carbonaceous-to-silicate dust in galaxies. Graphite particles
posses a strong absorption
feature at 2175~\AA\ seen in the extinction curve towards many stars in the
Milky Way (MW) and nearby
galaxies. The ratio of the 2175~\AA-to-continuum extinction can be used to
estimate the relative
abundance of graphite-to-silicate dust in the interstellar medium towards
the extincted source. The
right panel of Figure~1 is a schematic representation of the observed
average extinction in several
local galaxies. The Figure shows an obvious trend of increasing strength of
the 2175~\AA\ feature from
the SMC, LMC, the MW, and M51. The numbers next to each curve represent the
average metallicity in these
galaxies in solar units. The Figure suggests a trend of increasing
graphite-to-silicate dust ratio with
metallicity. The trend may suggest that details of the
galactic star formation history and stellar initial mass function play an
important role in
determining the dust composition and abundance, and hence the opacity, in
galaxies.

\begin{figure}
\plotone{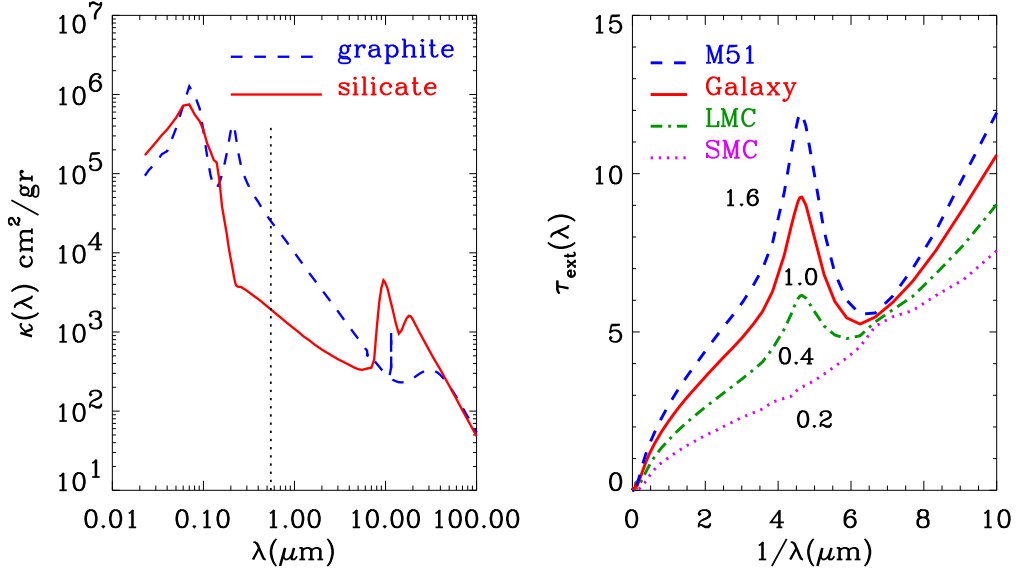}
\caption{Left panel: the mass absorption coefficient of graphite and
silicate grains. The dotted line gives the V-band position. Right panel: the
average interstellar extinction in several local galaxies. Curves are
labeled by the metallicity of the galaxies, normalized to that of the Milky
Way.}
\end{figure}

\section{Relative Distribution of the Dust Compared to the Sources}
Given a dust abundance and composition, the most important parameter
that determines the absorption efficiency of starlight and the
spectrum of the reradiated infrared emission is the proximity of the
dust to the radiation sources. One can distinguish between four
different dust environments characterized by the spatial relation
between the dust and the radiation sources: the circumstellar
environment, consisting of dust that had recently formed out of the
stellar ejecta and that is heated by the underlying stellar radiation
field; the interstellar environment, consisting of dust residing in
the ISM and heated by ionizing stars and/or the general interstellar
radiation field; the AGN environment, consisting of a dusty torus,
heated by the emission from the accretion disk; and the intergalactic
environment, consisting of dust that has been expelled from galaxies
and heated by the general diffuse background radiation.

\subsection{Circumstellar Dust}

Circumstellar dust dominates the mid--IR emission from intermediate
age stellar populations (Bressan, Granato, \& Silva 1998), and may be
an important source of thermal mid--IR emission in elliptical galaxies
(Knapp, Gunn, \& Wynn--Williams 1992). However, ISOCAM 4 and 15~$\mu$m
images of some elliptical galaxies reveal that the morphology of the
15~$\mu$m dust emission component is significantly different from the
4~$\mu$m stellar emission component (Madden, Vigroux, \& Sauvage
1999), suggesting an interstellar origin instead. Dust around young
stellar objects will also have a very high efficiency for converting
the radiation from the underlying object into infrared
emission. However, the lifetime of this embedded phase is very short
compared to the main sequence lifetime of these objects. Circumstellar
dust therefore plays a minor role in the redistribution of energy in
galaxies.

\subsection{Interstellar Dust and the Infrared Spectrum of Galaxies}

Most of the processing of galactic starlight is done by interstellar
dust particles that reside in the different phases of the ISM. In
fact, the infrared SEDs of all galaxies in the local universe can be
constructed from a linear combination of several distinct emission
components representing the different phases of the ISM: (1) a cirrus
component, representing the emission from dust and carriers of the
solid state infrared bands at 3.3, 6.2, 7.7, 8.6, 11.3, and
12.7~$\mu$m, both residing in the diffuse atomic phase of the ISM and
heated by the general interstellar radiation field; (2) a cold dust
component, representing the emission from dust residing in molecular
clouds, and heated by an attenuated interstellar radiation field; and
(3) an H~II or starburst emission component, representing the emission
from dust residing in H~II regions and heated by the ionizing
radiation field. An additional AGN component may be needed to
represent the spectra of some of the most luminous infrared
galaxies. Using this simple procedure with two or more emission
components, one can reproduce the fluxes and colors of {\it IRAS}
galaxies with luminosities ranging from normal ($L
\sim 10^{8.5}\ {\rm L}_{\odot}$) to the most luminous ($L \sim 10^{13}\
{\rm L}_{\odot}$) galaxies, and the observed trend of
increasing
$S(60\ \mu$m)/$S(100\ \mu$m) and decreasing $S(12\ \mu$m)/$S(25\ \mu$m)
flux ratios with increasing
infrared luminosity.

These empirical models offer a simple way of calculating the infrared
spectra of galaxies in the local universe (Malkan \& Stecker 1998;
Dwek et al.\ 1998; Guiderdoni et al.\ 1998; Rowan-Robinson \& Crawford
1989). However, to preserve the radiative energy balance of a galaxy,
this emission must equal the amount of starlight absorbed by the
dust. Calculating the opacity of galaxies poses a significant
challenge, since in addition to the microscopic dust properties, the
efficiency at which dust absorbs stellar photons depends on the dust
abundance, the clumpiness of the ISM, and the relative distribution of
stars and dust -- all of which are evolving quantities.

Various radiative transfer models attempting to represent this complex
reality were presented in the workshop on ``The Opacity of Spiral
Disks" (Davies \& Burstein 1995). Models that include a 2--phase ISM
consisting of molecular clouds and an intercloud component that
calculate not only the attenuation of starlight but the re-radiated IR
emission were developed, among others, by Silva et al.\ (1998),
V\'arosi \& Dwek (1999), and Misselt et al.\ (2000). The V\'arosi
\& Dwek model is analytic, and provides a very good approximation to the
Monte--Carlo model of Witt
\& Gordon (1996) that calculates the attenuation of radiation due to
absorption and multiple
scattering from clumpy spherical systems.
The models of Silva  et al., which include a clumpy interstellar medium, are
quite
successful in fitting the UV to submillimeter wavelength emission of galaxy
types ranging from
ellipticals and spirals to starbursts and interacting systems. The observed
SED of representative
galaxies is characterized by an increasing L$_{IR}$/L$_{opt}$ ratio along
the sequence from giant
ellipticals,  to spirals (NGC~6946), to
starbursts (M~82), to mergers (Arp~220). Silva et al.\ attribute this
trend to increasing infrared emission contributions from dust in giant
star-forming molecular clouds. In
fact, the infrared spectrum of Arp~220 is almost identical to that of an
H~II region. These models
provide a more physical approach to the construction of galaxy spectra,
using radiative transfer methods to calculate the contribution of the
various ISM dust components to the overall spectrum of the various types of
galaxies.

\subsection{Dusty Torii Around Active Galactic Nuclei}
The energy output from AGN represents the radiative energy
budget of the universe exclusive of the CMB and that released in
nuclear burning processes. The energy of an AGN is derived from the
release of gravitational energy associated with the accretion of
matter onto a central black hole (BH) located in the nucleus of a host
galaxy.  A significant fraction of this energy can be absorbed by the
dusty torus around the central BH. The energy output from AGN-dominated
galaxies is the sum of the reradiated thermal dust emission
and the non-thermal synchrotron emission. The overall spectral energy
distribution from these objects depends on the viewing angle. AGN that
are viewed face-on have a synchrotron dominated power-law spectrum,
whereas AGN that are viewed edge-on exhibit a thermal infrared excess
consisting of a hot ($\sim$ 50 K) component commonly attributed to
emission from the dusty torus, and a cooler ($\sim$ 20 K) component,
commonly attributed to reradiated stellar energy (Haas et al.\ 1998). In principle, the total bolometric contribution of AGN to the EBL could be comparable to the starlight contribution.  However,
direct observational correlation between X--ray and IR/submillimeter
sources suggests that AGN contribute only $\sim$ 10--20\% of CIB
intensity at 100 and 850~$\mu$m (Barger et al.\ 2000). Their
contribution at shorter wavelengths depends on the exact shape of the
IR spectrum and is therefore still uncertain.

\subsection{Intergalactic Dust}

The progressive dimming of the light output of Type Ia supernovae with
redshift has been taken as evidence that the mass density of the
universe is subcritical, requiring a cosmological constant for closure
(Perlmutter et al.\ 1999). An alternative scenario was suggested by
Aguirre (1999) who argued that intergalactic dust could produce the
same observational effect, alleviating the need to abandon the concept
of a flat universe with a zero cosmological constant. Such
intergalactic dust would be heated by the ambient intergalactic
radiation field and produce a truly diffuse infrared
background. Aguirre \& Haiman (2000) calculated the contribution of
such dust to the CIB provided it was sufficiently abundant to account
for the dimming of the distant supernovae. In particular, they found
that intergalactic dust would produce most of the CIB at
850~$\mu$m. However, sources detected by the SCUBA survey account for
over 50\% of the CIB at this wavelength, leaving little room for any
diffuse emission component.

\section{The IR Evolution of Galaxies and Models for the CIB}

In order to calculate the contribution of galaxies to the CIB, one
needs to know how their cumulative spectral energy density in the
universe has evolved with time. Various models have been put forward
to predict the CIB. These models can be grouped into four general
categories: backward evolution, forward evolution, semi analytical,
and cosmic chemical evolution models (see review by Hauser \& Dwek
2001). They differ in their degree of complexity, physical realism,
and ability to account for various observational constraints or to make
predictions. Backward evolution models are the simplest. They
extrapolate the spectral properties and/or the comoving number density
of local galaxies to higher redshifts using some parametric or
unphysical form for their evolution. The main disadvantage of these
models is that they are not constrained by the physical processes,
such as star and metal formation, or radiative transfer processes that
go on in the galaxies they represent.

Some of the shortcomings in backward evolution models are corrected in
forward evolution models. At the heart of these models is a spectral
evolution code which evolves the stellar populations and calculates
the stellar, gas, and metallicity content and SED of a galaxy as a
function of time. Initial conditions and model parameters are adjusted
to reproduce the observational properties of galaxies in the local
universe.  Models for the diffuse interstellar dust emission vary in
degree of complexity and physical input. As in the backward evolution
models, the IR emission is represented by a sum of two or more
components corresponding to the gas phases in which the dust resides
and the radiation field it is exposed to. The various dust emission
components are then evolved backwards in time in a manner that is
determined by the evolution of the various physical parameters that
determine their {\it present} intensity and spectral energy
distribution.

Detailed spectral evolution models that follow the evolution of the
dust composition and abundance, the galactic opacity, and the
UV-to-far infrared spectral energy distribution for various stellar
birthrate histories were constructed by Dwek (1998) and Dwek, Fioc, \&
V\'arosi (2000). The results of these calculations for spiral galaxies
are shown in Figure~2.

\begin{figure}
\plotone{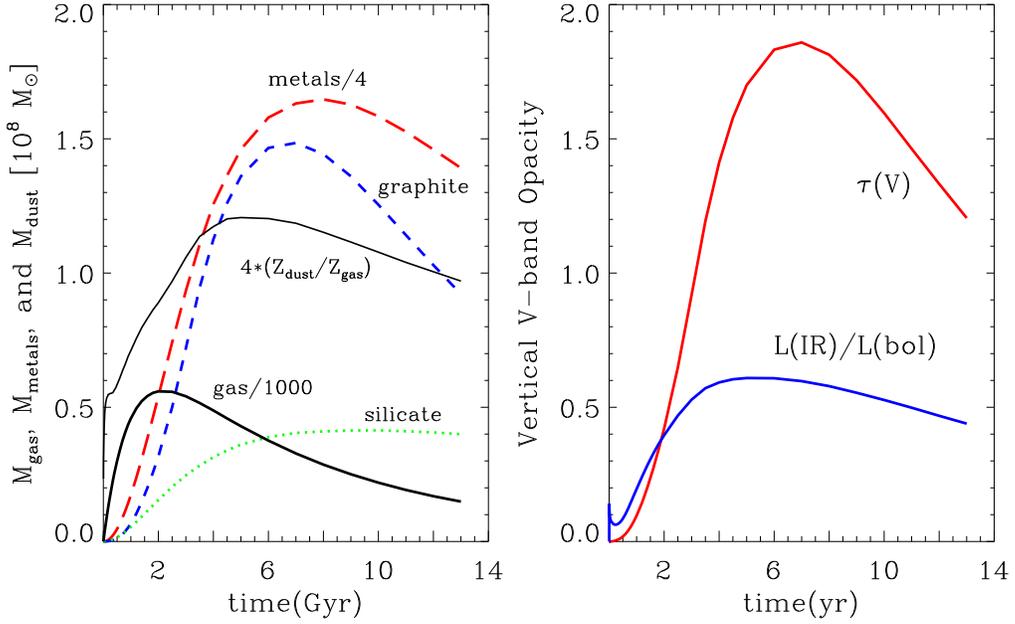}
\caption{The evolution of various observable quantities in spiral galaxies
as a function of time. See text for details.}
\end{figure}

The left panel of the Figure shows how the dust-to-metals as well as
the carbon-to-silicate mass ratios evolve with time. The evolutionary
trends reflect the temporal behavior of the evolution of the different
stellar sources (carbon stars, OH/IR stars, supernovae) that give rise
to the dust composition. The right panel depicts the evolution of the
V-band opacity perpendicular to the plane of the galaxy as a function
of time. The Figure shows that a maximum opacity is reached at an
epoch of $\sim$ 6~Gyr. Figure~3 shows the various emission components
contributing to the SED of a typical spiral galaxy at 12~Gyr (Fioc \&
Dwek 2000). They include the infrared emission from H~II regions and
diffuse H~I clouds. The contribution to the latter from PAH molecules,
carbon and silicate dust is explicitly shown in the Figure.

\begin{figure}
\plotone{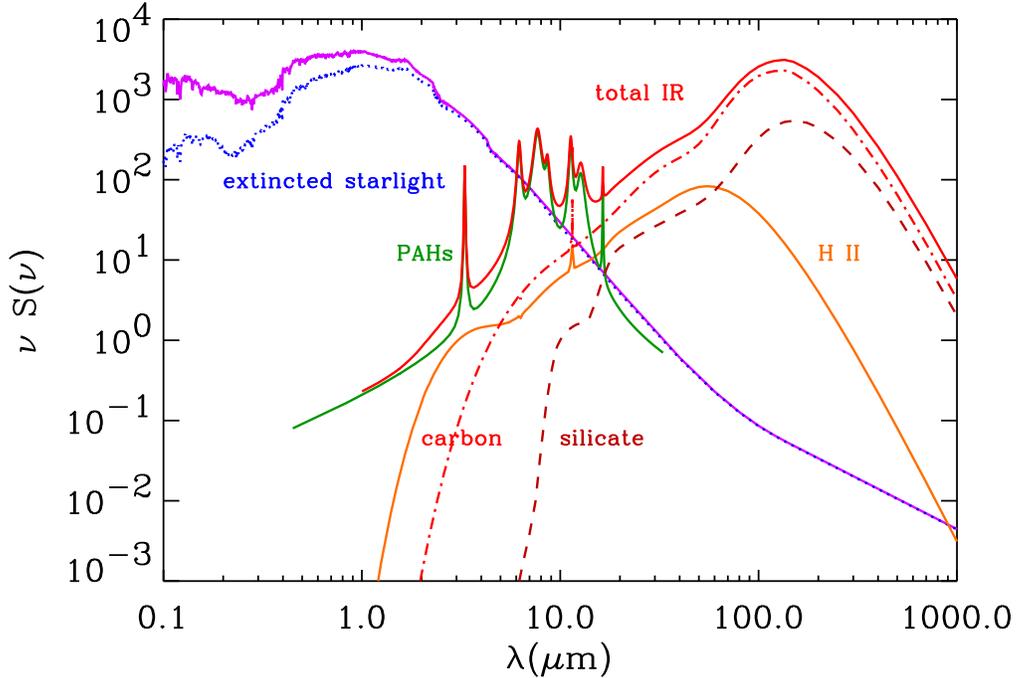}
\caption{The UV to far-infrared spectrum of a spiral galaxy, and the various
emission components contributing to the total spectrum.  The top curve
slanting down from upper left to lower right represents the unattenuated
starlight from the galaxy.
}
\end{figure}

Figure~4 examines the effect of the evolution of the dust composition and
abundance on the SED of spiral galaxies, by plotting the spectral ratio of
S$_{\nu}$(mw) to S$_{\nu}$(evol), versus wavelength for various epochs. The
spectrum S$_{\nu}$(mw) is the SED calculated under the assumption that the
dust-to-metal and the graphite-to-silicate mass ratios are constant, and
equal to their currently observed Milky Way values of 0.35 and 0.75,
respectively. The spectrum S$_{\nu}$(evol) includes the detailed evolution
of the metallicity dependent dust-to-gas and graphite-to-silicate dust mass
ratios, and galactic opacity as depicted in Figures 2 and 3. The Figures
show that non-evolving dust models over-predict the infrared emission at
early epochs, especially the mid-infrared emission bands. Both models relax
to the same spectrum at about 12~Gyr.

\begin{figure}
\plotone{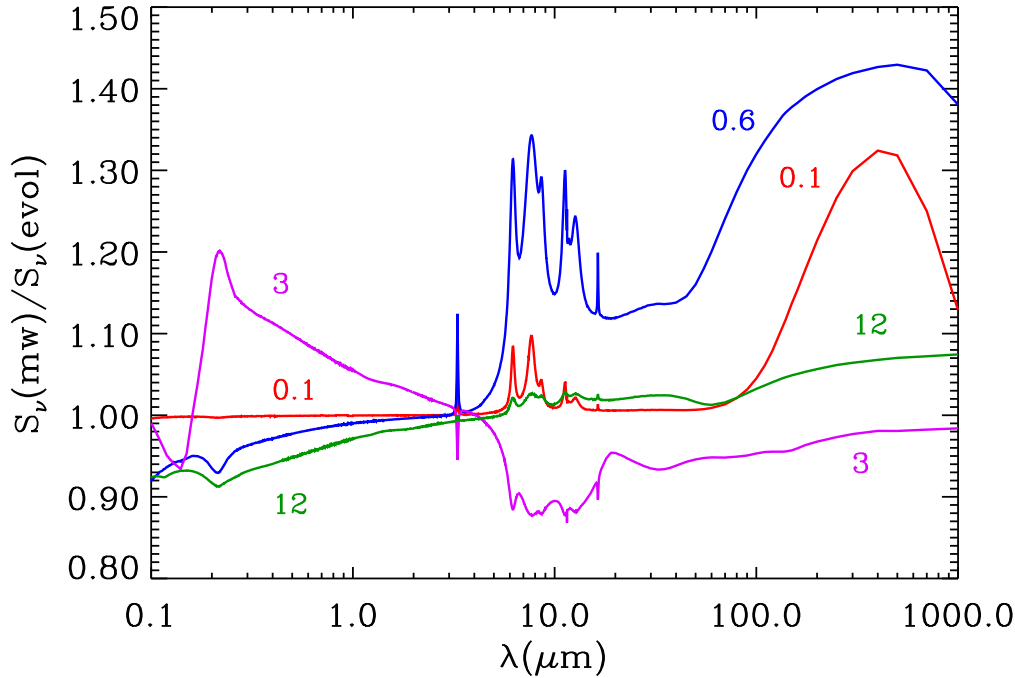}
\caption{The spectral ratio S$_{\nu}$(mw)/S$_{\nu}$(evol) versus wavelength
for various epochs in Gyr. See text for details.}
\end{figure}

In forward evolution models galaxies evolve quiescently, with no
allowance for any interaction or any stochastic changes in their star
formation rate or morphology. In particular these models fail to match
the SCUBA 850~$\mu$m galaxy number counts without the ad hoc inclusion
of a new population of ultraluminous infrared galaxies.  The SCUBA
observations of dust enshrouded galaxies at redshifts of $z \approx$ 2
-- 3, suggest a quick rise in galactic opacity. Figure~5 (based on the
calculations of Dwek 1998; V\'arosi
\& Dwek 1999; and Dwek, Fioc, \& V\'arosi 2000), shows the evolution of gas,
metals, and
dust mass in a pristine starburst (left panel) and the evolution of the
attenuation of
starlight (right panel) as a function of time. The Figure shows that
starbursts can become 
effectively opaque in the UV and optical in a mere 100 Myr.

\begin{figure}[ht]
\plotone{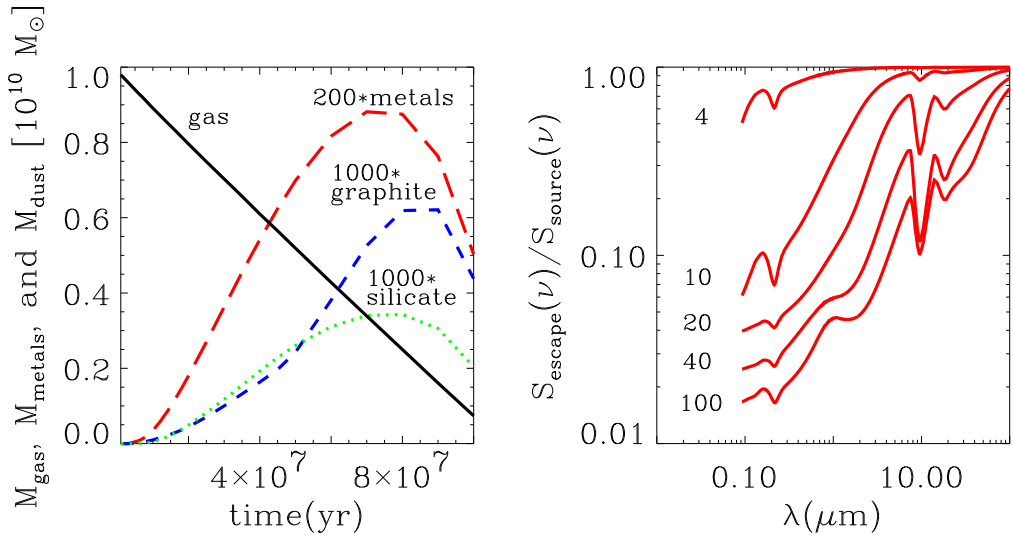}
\caption{The evolution of various parameters (left panel) and the fraction
of escaping radiation (right panel) from evolving pristine starbursts. The
curves in the right panel are marked by the starburst's age in Myr.}
\end{figure}

Most models for the CIB make similar predictions for the intensity and
spectral energy distribution of the CIB, regardless of their degree of
complexity or realism (Hauser \& Dwek 2001). This should not be
surprising, since the CIB is the cumulative sum of energy outputs in
the universe, and many details of the emission may be ``washed out''
in the summation process. The CIB is therefore not a strong
discriminator between models, especially since it is still not well
determined in the mid-infrared spectral region.

\section{Summary}

The very existence of the CIB provides conclusive evidence that we live in a
dusty universe. The CIB is
produced by the absorption and reemission of (mostly) starlight by
interstellar dust in normal and
active galaxies.

The CIB intensity and spectrum depend on a large number of microscopic
and global parameters that affect the optical properties of the dust,
its abundance and size distribution, the spatial distribution of the
dust relative to the radiation sources, and the clumpiness of the
interstellar medium in galaxies.  Some physical processes, such as the
stochastic heating of very small dust particles and macromolecules,
may be responsible for a significant fraction of the CIB in the mid-infrared.
In individual galaxies all these microscopic
and global properties evolve with time. For example, it is very likely
that the clumpiness in galaxies and the clump filling factor evolve in
a fashion that depends on the total mass of gas in the galaxy, the
supernova rate, and its metal content. Understanding the various
physical processes that determine why local galaxies are the way they
are, and how they evolved to their present conditions are crucial for
predicting their spectral appearance at various redshifts.

The CIB represents, however, the time integrated emission from
galaxies. There is therefore no unique way to determine the evolution
of the various microscopic and global parameters from studying the CIB
alone.

Evolutionary models show that massive star forming regions rapidly
become opaque at visible wavelengths.  Consequently, the total energy
released by these objects is mostly deposited in the CIB. The CIB can
therefore provide a useful integral constraint on the star formation
history of the universe. Different wavelength regions of the CIB
sample the contribution of galaxies at different redshifts. Studies of
the CIB can therefore provide important insights into the nature and
evolution of the various sources contributing to the CIB.

\acknowledgments
I thank Mike Hauser for his comments and input on some sections of
this paper, Ant Jones for his careful reading of and comments on
the manuscript, and Rick Arendt for assistance in preparing the Appendix of
the manuscript.  This work was supported by the NASA Astrophysical
Theory program OSS NRA 99-OSS-01.

\section*{Appendix: Probing the Cosmic Infrared Background with TeV
$\gamma$-ray Sources} 
Co-authored with Okkie C. de Jager\\
{\it Potchefstroom University, Potchefstroom 2520, South Africa}\\

The TeV spectrum of $\gamma$-ray sources can be attenuated by pair producing
$\gamma$-$\gamma$ interactions with the extragalactic background light
(EBL). Using recent detections of the EBL in the UV-optical (UVO) and
far-infrared regions, we set limits on the cosmic infrared background (CIB)
intensity in the $\sim$5 to 60~$\mu$m region with minimal assumptions on the
intrinsic Mrk 501 source spectrum. The results are shown in the four panels
below. The upper left panel depicts the TeV spectra of Mrk~501 and Mrk~421. The
panel to its right depicts the constraints used to construct possible EBL
spectra. These include 3 possible UVO spectra, 100 (5$\times$5$\times$4
intensity combinations at 6, 30, and 60~$\mu$m), and a fixed far-IR
spectrum, for a total of 300 EBL spectra. The intrinsic Mrk 501 source
spectra, corrected for attenuation by all EBL combinations, are shown in the
lower left panel. Most source spectra exhibit an unphysical rise at $>$ 10
TeV energies. However, some spectra remain flat in E$^2$dN/dE, as shown for select cases in the lower right
panel. Two source spectra are shown for each choice of UVO spectrum. The
upper curve depicts an unphysical spectral behavior, whereas the lower one
corresponds to a dN/dE $\propto$ E$^{-2}$ power law. Preliminary upper
limits on the CIB, derived by excluding EBL spectra giving rise to
unphysical source spectra, are 5 nW m$^{-2}$ sr$^{-1}$ in the 6 - 30~$\mu$m
interval and $<$ 10~\mbox{nW\,m$^{-2}$\,sr$^{-1}$} at 60~$\mu$m, for H$_0$ = 
\mbox{70~km\,s$^{-1}$\,Mpc$^{-1}$.}

\begin{center}
\begin{tabular}{cc}
\includegraphics[scale=0.35]{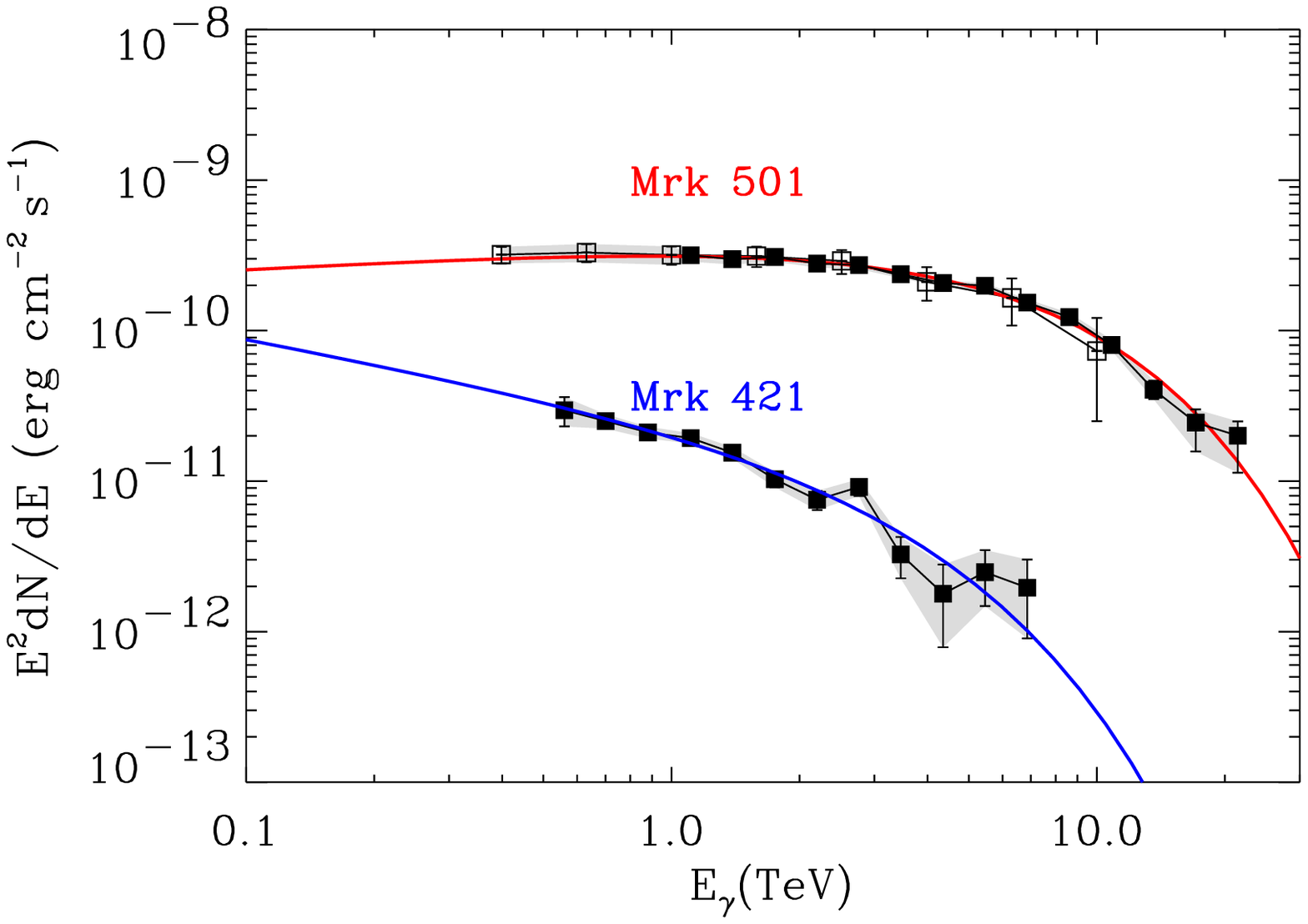}&
\includegraphics[scale=0.35]{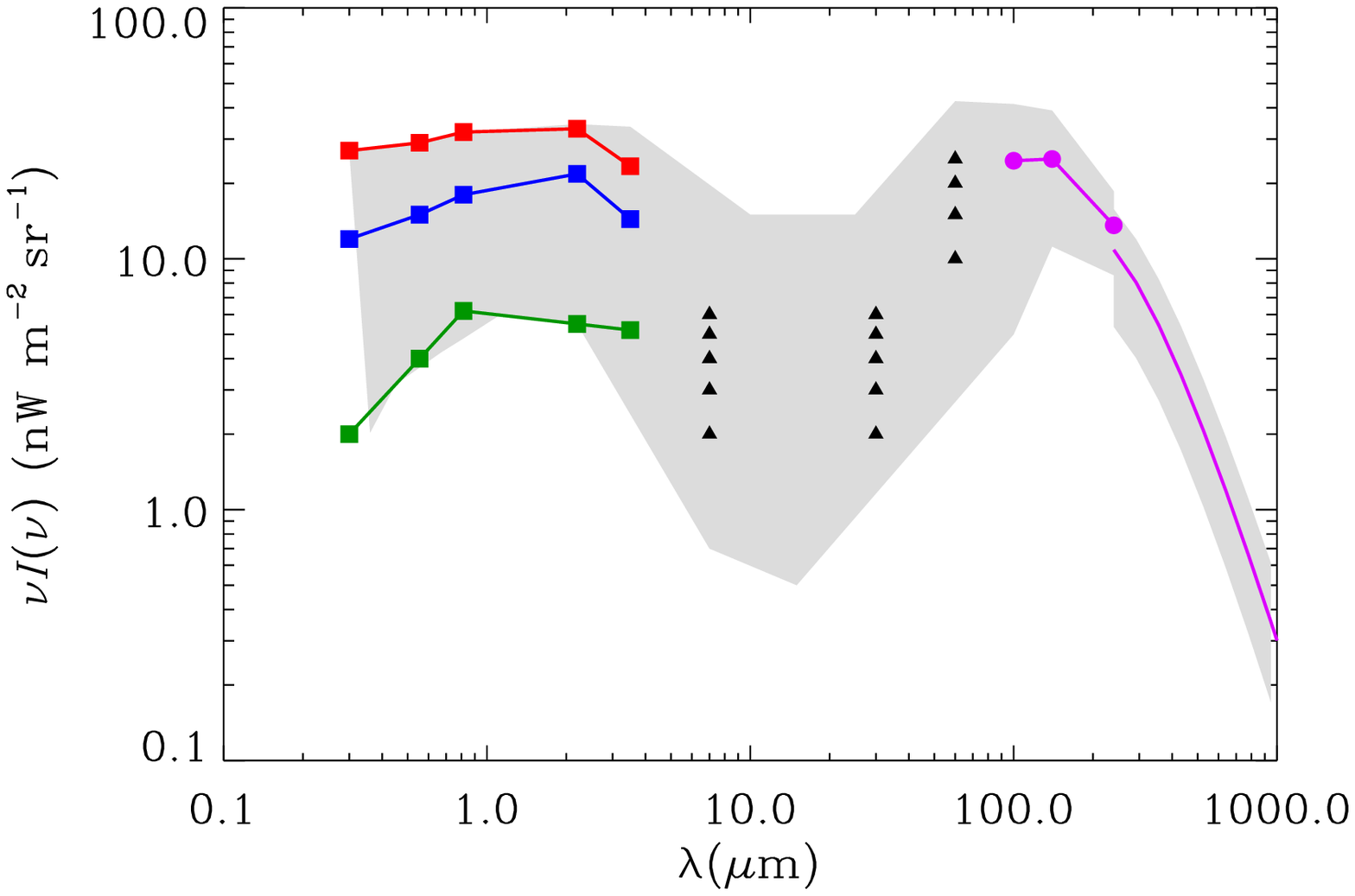}\\
\includegraphics[scale=0.35]{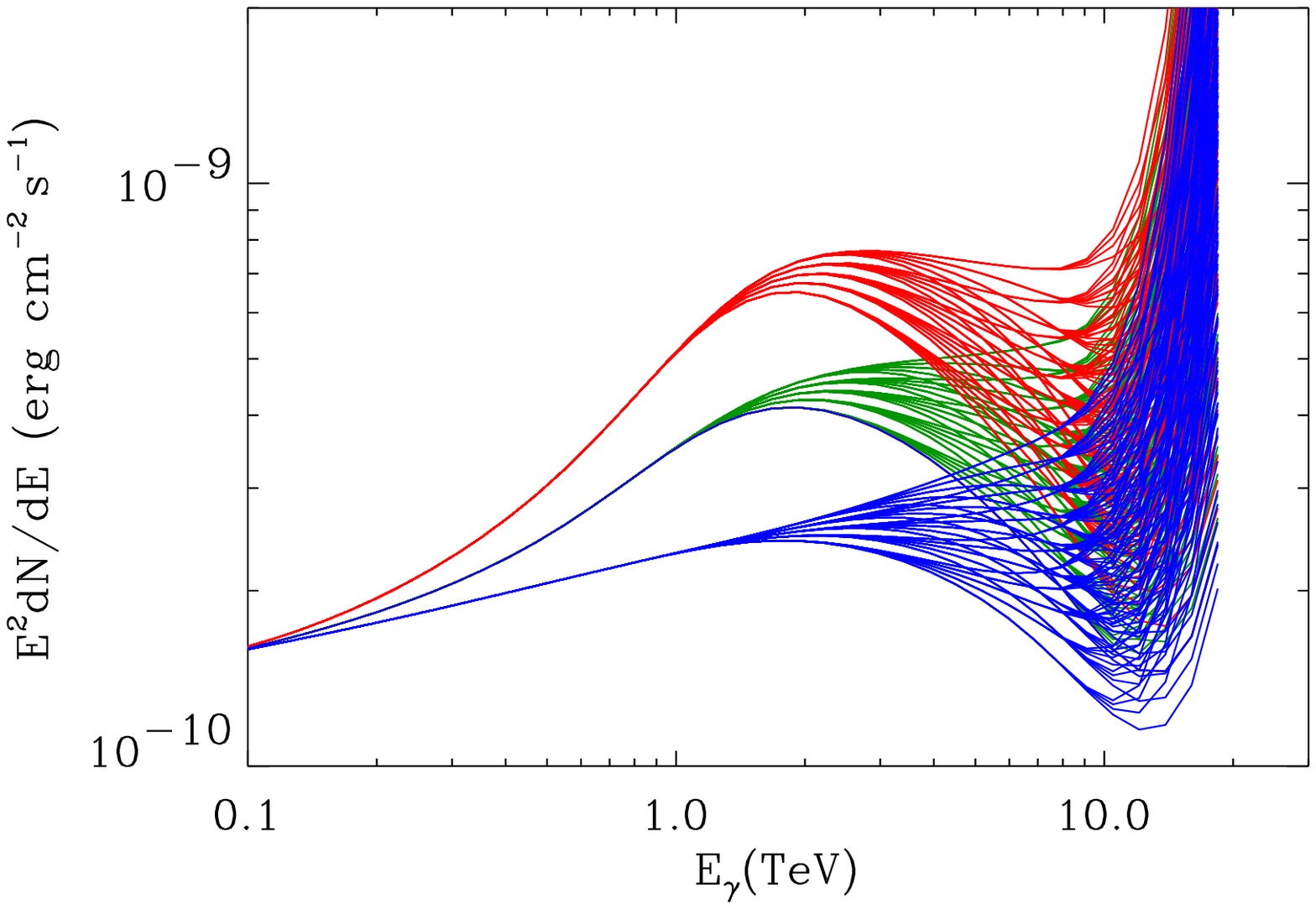}&
\includegraphics[scale=0.35]{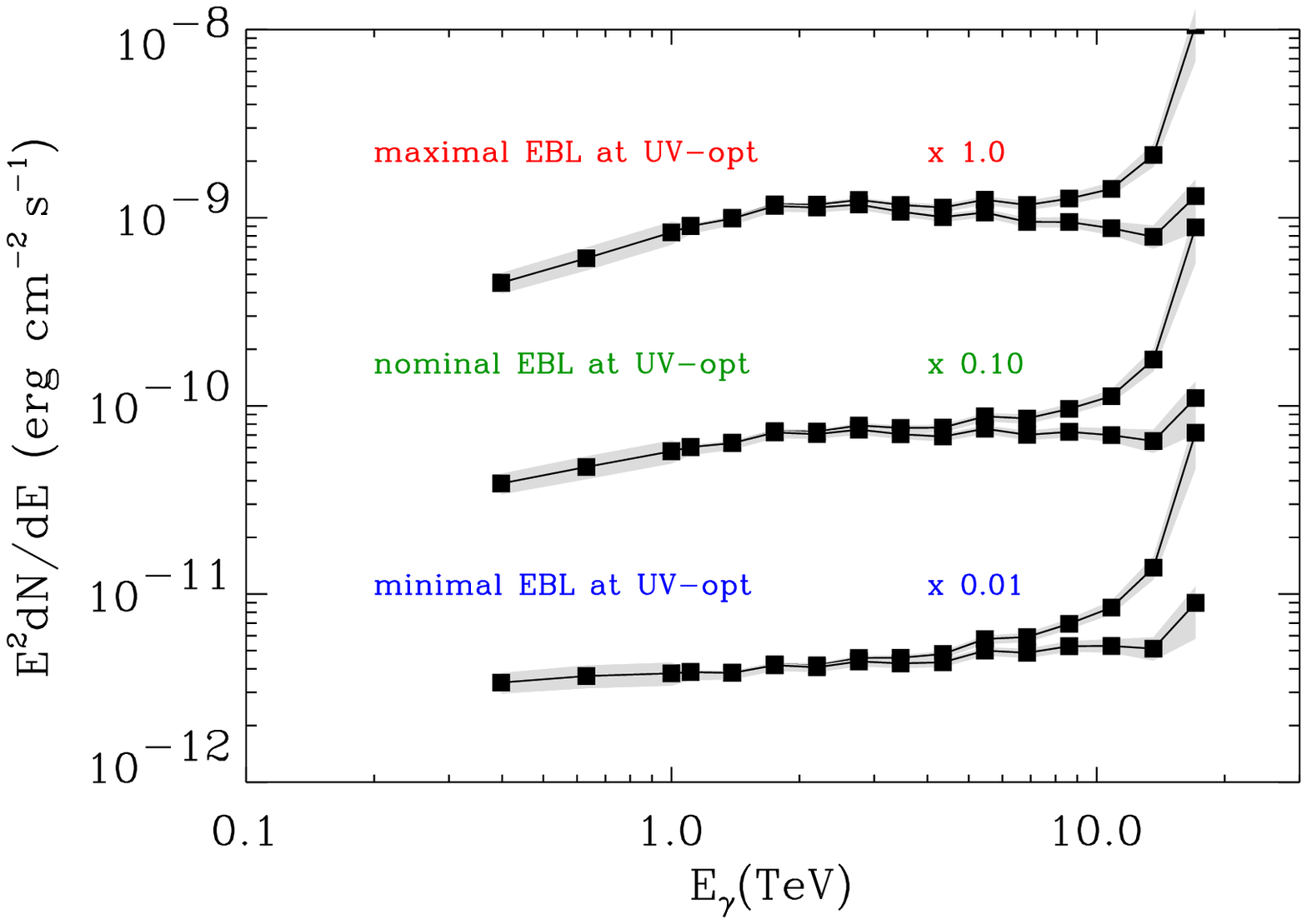}\\
\end{tabular}
\end{center}

\vspace{5mm}
\centerline{Discussion}
\vspace{5mm}

Mike Hauser:  The shaded region in your figure comparing models with CIB
measurements is the same as the $\pm 2\sigma$ region in my talk on the
observations of the CIB, earlier in this Symposium.

Eli Dwek:  Right.

\end{document}